\RequirePackage[l2tabu,orthodox]{nag}
\documentclass[envcountsect,envcountsame]{eptcs}
\providecommand{\event}{QPL 2017}
\providecommand{\authorrunning}{Dunne}
\providecommand{\titlerunning}{Spectral Presheaves, Kochen--Specker 
Contextuality, and Quantale--Valued Relations}

\usepackage[T1]{fontenc}
\usepackage{fixltx2e}
\usepackage{microtype} 
\usepackage{xspace}
\usepackage{multicol}
\usepackage{eurosym}

\makeatletter
\newcommand\etc{etc\@ifnextchar.{}{.\@}\xspace}
\makeatother

\usepackage{bbm}
\usepackage{graphicx}

\newcommand{\inlinegraphic}[2]{
  \dimendef\grafheight=255\dimendef\grafvshift=254
  \grafheight=#1
  \grafvshift=-0.5\grafheight
  \advance\grafvshift by 0.5ex
  \raisebox{\grafvshift}{\includegraphics[height=\grafheight]{images/#2}\xspace}
}

\newcommand{\ninlinegraphic}[2][1.0]{
  \dimendef\grafheight=255\dimendef\grafvshift=254
  \setbox0 = \hbox{\scalebox{#1}{\includegraphics{images/#2}}}
  \grafheight=\the\ht0
  \grafvshift=-0.5\grafheight
  \advance\grafvshift by 0.5ex
  \raisebox{\grafvshift}{\includegraphics[height=\grafheight]{images/#2}\xspace}
}

\newcommand{\QED}{\ensuremath{\null\hfill\square}}

\newcommand{\Alg}{\ensuremath{\textnormal{-\textbf{Alg}}}}
\newcommand{\Algvn}{\ensuremath{\Alg_{\textnormal{vN}}}}

\newcommand{\Spec}{\ensuremath{\textnormal{\text{Spec}}}}
\newcommand{\supp}{\ensuremath{\textnormal{\text{supp}}}}
\newcommand{\cosupp}{\ensuremath{\textnormal{\text{cosupp}}}}
\newcommand{\GSpec}{\ensuremath{\Spec_\textnormal{\text{G}}}}
\newcommand{\PSpec}{\ensuremath{\Spec_\textnormal{\text{P}}}}

\newcommand{\Hom}{\ensuremath{\textnormal{\text{Hom}}}}

\newcommand{\AlgA}{\ensuremath{\textnormal{\textbf{A}}}}

\newcommand{\AlgE}{\ensuremath{\textnormal{\textbf{E}}}}

\newcommand{\AlgB}{\ensuremath{\textnormal{\textbf{B}}}}
\newcommand{\AlgBool}{\ensuremath{\textnormal{\textbf{2}}}}

\newcommand{\Quant}{\ensuremath{(Q,\leq,\bigvee,\bot, \cdot, 1_Q)}}

\newcommand{\Set}{\ensuremath{\textnormal{\textbf{Set}}}}

\newcommand{\CTop}{\ensuremath{\textnormal{\textbf{Top}}}}
\newcommand{\RelQ}{\ensuremath{\Rel_{Q}}}

\newcommand{\RelB}{\ensuremath{\Rel_{\textbf{2}}}}
\newcommand{\Rel}{\ensuremath{\textnormal{\textbf{Rel}}}}
\newcommand{\FinRel}{\ensuremath{\textnormal{\textbf{FRel}}}}
\newcommand{\Hilb}{\ensuremath{\textnormal{\textbf{Hilb}}}}

\usepackage{latexsym}
\usepackage{amssymb}
\usepackage{amsmath} % can collide with some journals
\usepackage{stmaryrd}
\usepackage{mathtools}
\usepackage{amsthm}
\newtheorem{theorem}{Theorem}[section]

\newtheorem{lemma}[theorem]{Lemma}

\theoremstyle{definition}\newtheorem{example}[theorem]{Example}
\theoremstyle{definition}
\theoremstyle{definition}\newtheorem{definition}[theorem]{Definition}
\theoremstyle{definition}
\theoremstyle{definition}\newtheorem{remark}[theorem]{Remark}
\theoremstyle{definition}

\usepackage{diagrams} % remember to give shout-out to Paul Taylor
\newarrow{Equals}{}{=}{}{=}{}

\ifx\numericids\undefined
\newcommand{\id}[1]{\ensuremath{\mathrm{id}_{#1}}}
\else
\newcommand{\id}[1]{\ensuremath{1_{#1}}}
\fi

\newcommand{\catA}{\ensuremath{\mathcal{A}}\xspace}

\newcommand{\catC}{\ensuremath{\mathcal{C}}\xspace}

\newcommand{\op}{\ensuremath{^{\mathrm{op}}}\xspace}

\newcommand{\dg}{\ensuremath{\dag}}

\usepackage{hyperref}

\usepackage{mathrsfs} 

\usepackage{tikz}
\usetikzlibrary{arrows}

\usepackage{amsopn}

\begin{document}

\title{Spectral Presheaves, Kochen--Specker 
Contextuality, and Quantale--Valued Relations}
\author{Kevin Dunne
\institute{University of Strathclyde, Glasgow, Scotland.
\email{kevin.dunne@strath.ac.uk}}}
%\date{\today}

% For article.cls this goes after begin{document} 
% other classes may be differ
% \input{abstract.tex}

%\maketitle

\maketitle

%%% Local Variables: 
%%% mode: latex
%%% TeX-master: t
%%% End: 

\begin{abstract}
\emph{In the topos approach to quantum theory of Doering and Isham the 
Kochen--Specker Theorem, which asserts the contextual nature of quantum theory, 
can be reformulated in terms of the  global sections of a presheaf characterised 
by the Gelfand spectrum of a commutative $C^*$--algebra. In 
previous work we showed how this topos perspective can be generalised to a
class of categories typically studied within the monoidal approach to quantum 
theory of Abramsky and Coecke, and in particular how one can generalise the 
Gelfand 
spectrum. Here we study the Gelfand spectrum presheaf for categories 
of quantale--valued relations, and by considering its global 
sections we give a non--contextuality result for these categories. We 
also show that the Gelfand spectrum comes equipped with a 
topology which has a natural interpretation when thinking of these structures 
as representing physical theories. }
\end{abstract}

\section{Introduction}
\label{sec:Introduction}

The present work is part of an 
ongoing project \cite{Dunne2016:NewPerspective,Dunne2017:OnTheStructure} to 
marry conceptually the monoidal approach to quantum 
theory initiated by Abramsky and Coecke 
\cite{AbramskyCoecke2004:CategoricalSemantics}, 
and the topos approach 
to quantum theory initiated by Butterfield, Doering, and Isham
\cite{Doering2008:WhatIsAThing,IshamButterfield1998:AToposPerspective}. 
Both of these 
approaches to 
quantum theory are algebraic, in that they seek to represent some aspect of 
physical reality with algebraic structures. By taking the concept of 
a ``physical observable'' as a fixed point of reference we cast the difference 
between these 
approaches 
as internal vs. external algebraic perspectives; that is, algebras
\emph{internal} to a monoidal category $\catA$ vs. 
representations of algebras on $\catA$, a construction \emph{external} to 
$\catA$.  The topos approach to quantum theory
considers 
representations of commutative algebraic structures (for example 
$C^*$--algebras, or von Neumann algebras 
\cite{Conway2000:ACourseInOperatorTheory}) on $\Hilb$. The topos approach makes 
essential use of the fact that the sets $\Hom(H,H)$ in $\Hilb$ carry the 
structure of a $C^*$--algebra. In 
\cite{Dunne2016:NewPerspective} we 
showed that the categories considered in the monoidal approach have a similarly 
rich algebraic structure on their sets of endomorphisms $\Hom(A,A)$, thus 
allowing 
one to take the ``external perspective'' for any such $\catA$, and not just 
$\Hilb$.

There are various incarnations of the topos approach to quantum theory, here we 
follow a construction introduced in 
\cite{Doering2008:WhatIsAThing}, which is developed in 
\cite{Flori2013:Topos}. For 
a fixed Hilbert space $H$ one takes $\Hilb\Alg(H)$ to be the poset of 
commutative 
$C^*$--subalgebras of $\Hom(H,H)$ considered as a category, and
$\Hilb\Algvn(H)$ its subcategory whose objects are the commutative von 
Neumann $C^*$--subalgebras of $\Hom(H,H)$. We will briefly 
discuss a physical interpretation for this definition. Physical experiments have 
made clear that quantum mechanical 
systems are faithfully
represented by non--commutative $C^*$--algebras of the form 
$\Hom(H,H)$. What nature does not make clear however is 
how to 
interpret this algebraic structure. According to Bohr's 
interpretation of quantum theory \cite{Bohr1949:DiscussionWithEinstein}, 
although physical reality is by its nature quantum, as classical beings 
conducting 
experiments in our labs we only have 
access to the ``classical parts'' of a quantum system. Much of 
classical physics can be reduced to the study of commutative algebras; this 
approach is carefully constructed and motivated in 
\cite{Nestruev2003:SmoothManifoldsAndObservables} where the following picture is 
given:

\begin{align*}
\text{Physics lab}&  \qquad\qquad\to&&\text{Commutative unital} \\
& &&\mathbb{R}\text{--algebra } A \\
 \text{Measuring device}& \qquad\qquad\to&&\text{Element of the algebra } A\\
\text{State of the observed}& \qquad\qquad\to&& \text{Homomorphism of unital}  
\\
\text{physical system}& &&\mathbb{R} \text{--algebras } h:A \to \mathbb{R}\\
\text{Output of the}& \qquad\qquad\to&& \text{Value of this function } 
h(a), \\
\text{ measuring device}& && a \in A 
\end{align*}

\begin{center}Figure 1: Algebraic interpretation of classical physics
\end{center}

In \cite{Nestruev2003:SmoothManifoldsAndObservables} the author stresses that 
the choice of ground ring is somewhat unimportant to this construction and 
interpretation, however $\mathbb{R}$ is a reasonable choice given that in 
classical physics most of the quantities we want to measure, length, energy, 
time, etc., can be represented by real numbers. In quantum mechanics one 
traditionally takes scalar values in $\mathbb{C}$, but one can take any ring, 
or, as we will see, a semiring in its place and the physical 
interpretation of Figure 1. remains valid.

According to Bohr's interpretation of quantum theory, a quantum system 
represented by a non--commutative algebra $\Hom(H,H)$, can only be 
understood in terms of of its 
classical components, that is, the commutative subalgebras of 
$\Hom(H,H)$; in particular, we can only make observations on the classical 
subsystems. Hence the category $\Hilb\Algvn(H)$ is a collection of observable 
subsystems of a physical system, and we consider all of these subsystems at one 
by considering the category of 
presheaves $[\Hilb\Algvn(H)^{\op}, \,\Set]$, which is a topos. One presheaf of 
particular significance is the presheaf which is characterised by the 
\emph{Gelfand 
spectrum}. 
Recall the Gelfand spectrum of a commutative $C^*$--algebra $\AlgA$ is 
characterised by the set
$\GSpec(\AlgA) = \{ \ \rho:\AlgA \to \mathbb{C} \ | \ \rho \, \text{ a } \
C^*\text{--algebra homomorphism } \}
$ of \emph{characters}
which defines a functor
\[
\begin{tikzpicture}
\node(A) at (0,0) {$\Hilb\Algvn(H)^{\op}$};
\node(B) at (3,0) {$\Set$};
\draw[->](A) to node [above]{$\GSpec$}(B);
\end{tikzpicture}
\]
with the action on morphisms given by restriction. By Figure 1. 
we interpret 
this functor as assigning to each classical subsystem its set of possible 
states.

\begin{remark}\label{rem:DifferentSpectra}
The \emph{prime spectrum} $\PSpec(\AlgA)$ of a commutative $C^*$--algebra 
$\AlgA$ is defined to be the set of prime ideals of $\AlgA$, and is naturally 
isomorphic to the Gelfand spectrum. The correspondence comes from 
the fact that an ideal $J \subset \AlgA$ is prime if and only if it is the 
kernel of a character $\rho: \AlgA \to \mathbb{C}$. The prime spectrum is also 
equivalent to 
the \emph{maximal spectrum}, taken to be the collection of maximal ideals.
\end{remark}

In a presheaf category one can generalise the notion of elements of a set by 
considering the morphisms out of the terminal object. The terminal object $T: 
\catC^{\op}\to \Set$ in a presheaf category sends all objects to the singleton 
set $\{*\}$ and 
all morphisms to the identity $\id{}:\{* \} \to \{*\}$. A \emph{global section} 
(or \emph{global element}) of a presheaf $F: 
\catC^{\op} \to \Set$ is a natural 
transformation $\chi: T \rightarrow F$.

The Kochen--Specker 
theorem \cite{KochenSpecker1975:LogicalStructures} asserts the 
\emph{contextual} nature of quantum theory. The principle of 
\emph{non--contextuality} is that the outcome of a measurement should not depend 
on the \emph{context} in which that measurement is performed, that is, it should 
not depend on which other measurements are made simultaneously. Classical 
physics is typically formulated as non--contextual \cite[Chap. 
4]{Isham1995:LecturesOnQuantumTheory}. The Kochen--Specker theorem 
states that it is a feature of quantum 
theory that one can find collections of measurements for which the outcomes are 
context dependent. For a mathematical 
treatment of this theorem see in \cite[Chap. 
9]{Isham1995:LecturesOnQuantumTheory}.
The following 
theorem was first shown in \cite{Doering2008:WhatIsAThing} but here we 
present it as in \cite[Corollary 9.1]{Flori2013:Topos}.

\begin{theorem}\label{thm:IshamKS}
The Kochen-Specker theorem is equivalent to the statement that for a Hilbert 
space $H$ with $\dim (H)\geq 3$, the presheaf 
$\GSpec:\Hilb\Algvn(H)^{\op}\to \Set$
has no global sections.
\end{theorem}

The monoidal approach to quantum theory of Abramsky and Coecke 
\cite{AbramskyCoecke2004:CategoricalSemantics} is an entirely separate approach 
to quantum theory using very different mathematical structures. This approach 
begins with identifying the essential 
properties of the category $\Hilb$ which one needs to 
formulate concepts from quantum theory.

\begin{definition}
A $\dg$\emph{--category} consists of a category $\catA$ together with an 
identity on objects functor 
$\dg: \catA^{\op} 
\to \catA$ satisfying $\dg \circ \dg = \id{\catA}$. A
$\dg$\emph{--symmetric monoidal category} consists of a symmetric monoidal 
category $(\catA,\otimes,I)$ which is a $\dg$--category such that $\dg$ is a 
strict
monoidal functor and all of the 
symmetric monoidal structure isomorphisms satisfy $\lambda^{-1} = 
\lambda^{\dg}$.
\end{definition}

\begin{definition}
A category $\catA$ is said to have \emph{finite biproducts} if it has a zero 
object $0$, and if for all objects $X_1$ and $X_2$ there exists an object $X_1 
\oplus X_2$ which is both the coproduct and the product of $X_1$ and $X_2$.
If $\catA$ is a $\dg$--category and has finite biproducts such 
that the 
coprojections $\kappa_i : X_i \to X_1 \oplus X_2$ and projections $\pi_i : 
X_1 \oplus X_2 \to X_i$ are related by $\kappa_i^{\dg} = \pi_i$, then we say 
$\catA$ has \emph{finite} $\dg$\emph{--biproducts}.
\end{definition}

In a category with a zero object $0$, for every pair of objects $X$ and $Y$ 
we call the unique map $X \to 0 \to Y$ the \emph{zero--morphism}, which we 
denote $0_{X,Y}: X \to Y$, or simply $0:X \to Y$.

For a category with finite biproducts each hom-set $\Hom(X,Y)$ is 
equipped with a commutative monoid operation 
\cite[Lemma 18.3]{Mitchell1965:TheoryOfCategories} called \emph{biproduct 
convolution}, where for
$f,g:X \to Y$, we define $f+g:X \to Y$ by the composition
\begin{equation*}\label{eq:Enrich}
\begin{tikzpicture}[baseline=(current  bounding  box.center)]
\node(A) at (0,0) {$X$};
\node(B) at (1.9,0) {$X\oplus X$};
\node(C) at (4.1,0) {$Y\oplus Y$};
\node(D) at (6,0) {$Y$};
\draw[->](A) to node [above]{$\Delta$}(B);
\draw[->](B) to node [above]{$f \oplus g$}(C);
\draw[->](C) to node [above]{$\nabla$}(D);
\end{tikzpicture}
\end{equation*}
where the monoid unit is given by the zero--morphism $0_{X,Y}:X\to Y$.

Categories with finite $\dg$--biproducts admit a matrix 
calculus \cite[Chap. I. Sect. 17.]{Mitchell1965:TheoryOfCategories} 
characterised as follows. For 
$X= \bigoplus\limits_{j=1}^{n} 
X_j$ and $Y= \bigoplus\limits_{i=1}^{m} Y_i$ a morphism $f:X \to Y$ is 
determined 
completely by the morphisms $f_{i,j} : X_i \to Y_j$, and
morphism composition is given by matrix multiplication. If $f$ has matrix 
representation $f_{i,j}$ then $f^\dg$ has matrix representation $f_{j,i}^\dg$.

The category $\Hilb$ is the archetypal example of a category with these 
properties.
A notion of ``observable'' in quantum theory can be axiomatised in terms 
of the monoidal 
structure of the category of Hilbert spaces by Frobenius algebras
\cite{CoeckeEtAl2008:NewDescriptionOrthogonal} or $H^*$--algebras 
\cite{AbramskyHeunen2012:HAlgebras}, and hence we can consider any 
$\dg$--symmetric 
monoidal category as a categorical model for a toy theory of 
observables. For example, Spekkens Toy Theory \cite{Spekkens2007:Epistemic} 
 is a 
toy physical theory
exhibiting some quantum--like properties but which is given by a local hidden 
variable model. This theory can be modelled in the category of sets and 
relations $\Rel$ using Frobenius algebras to represent observables
\cite{CoeckeEdwards2008:ToyQuantum}.

The monoidal approach provides general framework in which a broad class of 
physical theories 
can be compared in a high--level but mathematically rigorous way. This is 
useful 
for exploring interdependencies of quantum or quantum--like phenomena, for 
example the many notions of non--locality and contextuality. In particular, in 
\cite{GogiosoZeng2015:MerminNonlocality} an abstract notion of 
Mermin--locality is formulated in the language of Frobenius algebras, and the 
category of finite sets and relations $\FinRel$ is shown to be Mermin--local.

In this work we present a completely abstract notion of Kochen--Specker 
contextuality and we show that categories of quantale--valued relations do not 
exhibit this form of contextuality. This is done using abstract 
Gelfand spectrum introduced in \cite{Dunne2016:NewPerspective}. In order to 
prove this non--contextuality result we define 
the \emph{prime spectrum} 
which we relate to the physical interpretation of Figure 1. by examining the 
topological structure which these spectra carry.

\section{The Spectrum and Kochen--Specker Contextuality}
\label{sec:StateSpaceTopos}

In this section we review a construction introduced 
in \cite{Dunne2016:NewPerspective}, and we introduce an abstract 
definition of 
Kochen--Specker contextuality. This is done using the language of semirings, 
semimodules \cite{Golan1992:TheoryOfSemirings} and semialgebras. 

\begin{definition}\label{def:Semiring}
A \emph{semiring} $(R,\cdot,1,+,0)$ consists of a set $R$ equipped with a 
commutative 
monoid operation $+: R \times R \to R$ with unit 
$0\in R$, and a monoid 
operation $\cdot : R \times R \to R$, with unit $1\in R$, such that $\cdot$ 
distributes over $+$ and such that  $s\cdot 0 = 0 \cdot s =0$ for all $s\in R$.

A semiring is called \emph{commutative} if $\cdot$ is commutative. A 
\emph{$*$-semiring}, or \emph{involutive semiring} is one equipped 
with an operation $*: R \to R$ which is an involution, a homomorphism for 
$(R,+,0)$, and satisfies $(s \cdot t)^* = t^* \cdot s^*$ and $1^* = 1$.
\end{definition}
As the 
notation 
suggests we will refer to the monoid operations of a semiring 
as \emph{addition} and \emph{multiplication} respectively. We say that a 
semiring $R$ is \emph{zero--divisor free (ZDF)} if for all 
$s,t \in R$ we have $s\cdot t = 0 $ implies $s=0 $ or $t=0$.
Many concepts associated with rings can be lifted directly to the level of 
semirings in the obvious way, for example homomorphisms, kernels and direct 
sums. However, some concepts have to be treated with care when generalising to 
semirings, for example \emph{ideals}. 

\begin{definition}
Let $R$ be a commutative semiring. A subset $ J\subset R$ is 
called an \emph{ideal} if it contains $0$, is closed under addition, and 
such that for all 
$s \in R$ and $a \in  J$, $as \in  J$. An ideal is called \emph{prime} if $st 
\in  J$ 
implies $s \in  J$ or $t \in  J$. A $k$--\emph{ideal} is an ideal $J$ such that 
if $a \in J$ and $a+b \in J$ then $b \in J$. A $k^*$--ideal of a $*$--semiring 
is a $k$--ideal closed under involutions.
\end{definition}

The $k$--ideals are to a semirings what normal 
subgroups are to a groups; 
they are the ideals which one can quotient by. For any ring considered as 
a semiring the ideals and $k$--ideals coincide.

\begin{definition}
Let $(R,\cdot ,1, +,0)$ be a commutative semiring, an  
$R$--\emph{semimodule} consists 
of a commutative 
monoid $+_M :M\times M \to M$, with unit $0_M$, together with a \emph{scalar 
multiplication} 
$\bullet: 
R \times M \to M$ such that for all $r,s \in R$ and $m,n \in M$:

\begin{multicols}{2}
\begin{enumerate}
 \item $s \bullet (m +_M n) = s\bullet m +_M s \bullet n$ ;
 \item $(r\cdot s) \bullet m = r\bullet ( s \bullet m)$ ;
 \item $(r+s) \bullet m = (r \bullet m) +_M (s \bullet m) $;
 \item $0 \bullet m = s \bullet 0_M = 0_M$;
 \item $1 \bullet m = m$.
\end{enumerate}
\end{multicols}
\end{definition}

\begin{definition}
An \emph{$R$--semialgebra} $(M,\cdot_M,1_M,+_M,0_M)$ consists of an 
$R$-semimodule $(M,+_M,0_M)$ 
equipped with a monoid operation $\cdot_M:M \times M \to M$, with unit $1_M$, 
such that 
$(M,\cdot_M,1_M,+_M,0_M)$ forms a semiring, and where scalar multiplication 
obeys 
$s 
\bullet (m \cdot_M n) = (s
\bullet m )\cdot_M n  =
 m \cdot_M (s \bullet n)$. An $R$--semialgebra is called 
\emph{commutative} if $\cdot_M$ is commutative.
\end{definition}

The ideals and $k$--ideals of a semialgebra are defined in the obvious way. 
Notice that every semiring $R$ is an $R$-semialgebra, where the scalar 
multiplication by $R$ is taken to be the usual multiplication in $R$. Non--zero 
elements $s,t$ of a semialgebra are \emph{orthogonal} if $s\cdot t = 0$. A
\emph{subunital idempotent} in a semialgebra is an idempotent element $p$ such 
that there is an orthogonal idempotent $q$ where $p+q = 1_M$. A 
\emph{primitive subunital idempotent} is a subununital idempotent $p$ such that 
there exists no non--trivial subunital idempotents $s$ and $t$ with $s+t = 
p$.

\begin{definition}\label{def:DaggerSemialgebra}
Let $R$ be a $*$--semiring. An \emph{$R^*$--semialgebra}
consists of an $R$--semialgebra $(M,\cdot_M,1_M,+_M,0_M)$, such that $M$ and 
$R$ have compatible involutions, that is, one that satisfies $(s \bullet 
m)^{*}=s^{*} \bullet m^{*}$. 
\end{definition}

A \emph{unital subsemialgebra} $i: N \hookrightarrow M$
of $M$ is a subset $N$ containing $0_M$ and $1_M$ closed under all the 
algebraic 
operations. A \emph{subsemialgebra} $N\subset M$ is a subset $N$ containing 
$0_M$ which is closed 
under multiplication and which is a semialgebra in its own right, though may 
have a different unit from $M$.
A 
\emph{(unital) $*$--subsemialgebra} of a $*$--semialgebra is a 
(unital) subsemialgebra closed 
under taking involutions.

An $R$--semialgebra is said to be \emph{indecomposable} if it cannot be 
expressed as a 
non--trivial direct sum of $R$--semialgebras. An $R$--semialgebra is 
\emph{completely 
decomposable} if it is isomorphic to the direct sum of its indecomposable 
subsemialgebras.

The following two results are shown in detail in 
\cite{Dunne2016:NewPerspective}.
\begin{theorem}
For a locally small $\dg$--symmetric monoidal category $(\catA, \otimes, I)$ 
with finite $\dg$--biproducts 
the 
set $S=\Hom(I,I)$ is a commutative $*$--semiring.
\end{theorem}

Biproduct convolution gives us the additive operation on $S$ while 
morphism 
composition gives us the multiplicative operation,  and the functor $\dg$ 
provides the involution. It is shown in 
\cite{KellyLaplaza1980:CoherenceForCompact} that this multiplicative operation 
is commutative.

\begin{theorem}
Let  $(\catA, \otimes, I)$ be a locally small $\dg$--symmetric monoidal 
category and let $S=\Hom(I,I)$. For 
any pair of objects the set $\Hom(X,Y)$ is an $S$--semimodule, and
 $\Hom(X,X)$ is a $S^*$--semialgebra.
\end{theorem}
Addition on the set $\Hom(X,Y)$ is given by biproduct convolution. For a 
morphism $f:X \to Y$ scalar multiplication $s \bullet f$ for $s:I \to I$ is 
defined
\[
\begin{tikzpicture}
\node(A) at (0,0) {$X$};
\node(B) at (1.8,0) {$X\otimes I$};
\node(C) at (4.2,0) {$Y \otimes I$};
\node(D) at (6,0) {$Y$};
\draw[->] (A) to node [above]{$\sim$}(B);
\draw[->] (B) to node [above]{$f \otimes s$}(C);
\draw[->] (C) to node [above]{$\sim$}(D);
\end{tikzpicture}
\]
which gives a semimodule structure \cite{Heunen2008:SemimoduleEnrichment}. 
For $\Hom(X,X)$ multiplication is given by morphism composition and the 
functor $\dg$ provides the involution.

\begin{definition}
For $(\catA, \otimes, I)$ a locally small $\dg$--symmetric monoidal category 
and 
$X$ and object, we define the category $\catA\Alg(X)$ to be the category with 
objects commutative unital $S^*$--subsemialgebras of $\Hom(X,X)$, and arrows 
inclusion 
of unital subsemialgebras.
\end{definition}

The for any subset of $B\subset \Hom(X,X)$ the set $B' = \{\ f:X \to X \ | \ 
f\circ g= g \circ f \text{ for all } g\in B \ \}$ is called the 
\emph{commutant} of $B$ \cite[Sect. 
12]{Conway2000:ACourseInOperatorTheory}. We 
define the full subcategory of 
\emph{von Neumann $S^*$--subsemialgebras}
\[
\begin{tikzpicture}
\node(A) at (0,0) {$\catA\Algvn(X)$};
\node(B) at (3,0) {$\catA\Alg(X)$};
\draw[right hook->](A) to node [above]{}(B);
\end{tikzpicture}
\]
to have objects those $S^*$--subsemialgebras $\AlgA$ which satisfy $\AlgA = 
\AlgA''$.

\begin{example}
If we take $(\catA, \otimes,I)$ to be $(\Hilb,\otimes, I)$ then the category 
$\Hilb\Algvn(H)$ is precisely the category considered in the topos approach
\cite{Doering2008:WhatIsAThing,Flori2013:Topos}.
\end{example}

\begin{remark}\label{rem:FrobeniusLifts}
In \cite{Dunne2017:OnTheStructure} we showed that any special commutative 
unital 
Frobenius algebra, and any (possibly non--unital) commutative $H^*$--algebra 
$\mu:A \otimes A \to A$ in 
$\catA$ generates an object 
$\AlgA$ in $\catA\Algvn(A)$. Furthermore, there is a natural
correspondence between 
the set--like elements of the internal algebra and the Gelfand spectrum of the 
semialgebra it generates. Hence the notion observable in the 
monoidal 
approach naturally lifts to the notion of observable
in our
generalised topos approach.
\end{remark}

We can
generalise the 
spectrum of a commutative $C^*$--algebra to an $S^*$--semialgebra 
\cite{Dunne2016:NewPerspective}.

\begin{definition}
Let $(\catA, \otimes, I)$ be a locally small $\dg$--symmetric monoidal category 
with finite 
$\dg$--biproducts, and $X$ an object. The \emph{Gelfand spectrum} 
for $X$ is the presheaf
\[
\begin{tikzpicture}
\node(A) at (0,0) {$\catA\Algvn(X)^{\op}$};
\node(B) at (3,0) {$\Set$};
\draw[->](A) to node [above]{$\GSpec$}(B);
\end{tikzpicture}
\]
defined on objects
$
\GSpec(\AlgA) = \{ \ \rho:\AlgA \to S \ | \ \rho \text{ an } 
S^*\text{--semialgebra homomorphism } \}
$ to be the set of 
\emph{characters}
while the action on morphism is defined in the obvious way by restriction.
\end{definition}

\begin{definition}
Let $(\catA,\otimes, I)$ be a locally small $\dg$--symmetric monoidal category 
with finite 
$\dg$--biproducts, and $X$ an object. The \emph{prime spectrum} 
for $X$ is the presheaf
\[
\begin{tikzpicture}
\node(A) at (0,0) {$\catA\Algvn(X)^{\op}$};
\node(B) at (3,0) {$\Set$};
\draw[->](A) to node [above]{$\PSpec$}(B);
\end{tikzpicture}
\]
defined on objects $\PSpec(\AlgA) = \{ \ J \subset \AlgA \ | \ J \text{ a prime 
$k^*$--ideal} 
\}$
while for $i:\AlgA 
\hookrightarrow \AlgB$ the action on morphisms is given by $\tilde{i}: 
\PSpec(\AlgB) \to 
\PSpec(\AlgA)$ 
is defined $\tilde{i}(K) = \{ \ x \in 
\AlgA \ | \ i(x) \in K \ \}$.
\end{definition}
To see that $\tilde{i}(K)$ is a prime $k^*$--ideal one can see the proof of a 
similar 
statement \cite[Proposition 6.13]{Golan1992:TheoryOfSemirings}.
\begin{remark}
One can also define a functor which assigns to each $\AlgA$ the collection of 
all prime ideals, not just the prime $k^*$--ideals 
\cite[Chap. 6]{Golan1992:TheoryOfSemirings}, although for the purposes of this 
work $k^*$--ideals are a more natural choice. One can define the maximal 
spectrum 
for an 
arbitrary semialgebra or semiring, although this fails to be functorial in 
general, \cite[Chap 2. Sect. 5]{Smith2014:AlgebraicGeometry}.
\end{remark}

We have already discussed that for $\Hilb$ the prime spectrum and Gelfand 
spectrum coincide. In \cite{Dunne2016:NewPerspective} we 
showed that the same is true for the category of sets and relations $\Rel$, 
although we will 
see in Example \ref{example:1} that this 
is not the case in general. 

The Gelfand spectrum presheaf formulation of the 
Kochen--Specker theorem justifies the 
following definition.

\begin{definition}\label{def:KScontextual}
Let $\catA$ be a locally small $\dg$--symmetric monoidal category with finite 
$\dg$--biproducts. An object $X$ in $\catA$ is said to be \emph{Kochen--Specker 
contextual} if the presheaf $\GSpec$ on $\catA\Algvn(X)$
has no global sections. We say $X$ is \emph{Kochen--Specker non--contextual} 
if $\GSpec$ 
does admit a global section.
\end{definition}

Such a global section, if it exists, will pick out an element 
$\chi_\AlgA : \{*\}\to \GSpec(\AlgA)$ from each spectrum, i.e. according to 
Figure 1. it would specify a state 
from 
each ``classical subsystem'' $\AlgA$. Naturality ensures that these choices of 
states are consistent with measurement outcomes irrespective of which subsystem 
 -- that is, which ``context'' -- the measurement
appears in.

There are more general formulations of contextuality and non--locality using 
the 
language 
of sheaves and presheaves
\cite{AbramskyBrandenburger2011:UnifiedSheafTheoretic,
AbramskyEtAl2015:ContextualityCohomology}.
Future work will show how the 
categories $\catA\Algvn(X)$ naturally generate \emph{empirical 
models} which can be examined within the framework of 
\cite{AbramskyEtAl2015:ContextualityCohomology}, and how the contextual nature 
of those empirical models is related to the existence of global sections for 
the 
corresponding Gelfand spectrum. This connection, together with Remark 
\ref{rem:FrobeniusLifts} will give us a means of applying the techniques of 
\cite{AbramskyEtAl2015:ContextualityCohomology}, for example sheaf cohomology, 
to the Frobenius algebras in an 
arbitrary $\catA$.

\section{Quantale--Valued Relations}
\label{sec:Quantales}

We now turn our attention to a class of categories for which we will prove a 
non--contextuality result, namely quantale--valued relations over a fixed 
quantale $Q$. A standard reference for quantales is 
\cite{Rosenthal1990:Quantales}.

\begin{definition}
A \emph{quantale} $(Q, \bigvee, \cdot, 1_Q)$ is a 
complete join--semilattice $(Q,\bigvee)$ equipped with a monoid 
operation $\cdot : Q\times Q \to Q$ with unit $1_Q$ such that for any $x \in Q$ 
and  
$P \subset Q$ 
\[
x \cdot (\bigvee\limits_{y \in P} y) = \bigvee\limits_{y \in P}(x \cdot 
y)\qquad \text{and}\qquad (\bigvee\limits_{y \in P} y)\cdot x = 
\bigvee\limits_{y \in P}(y \cdot x)
\]
An \emph{involutive quantale} in one equipped with an involution 
map $*:Q \to Q$ which is a semilattice homomorphism which is an involution 
$(x^*)^* = x$ satisfying $(x\cdot y)^* = y^*\cdot x^*$ and $1_Q^* = 1_Q$.
A \emph{commutative quantale} is one for which the monoid operation is 
commutative. A \emph{subquantale} is a subset of $Q$ closed under all joins and 
the 
monoid operation and containing $1_Q$.
\end{definition}

We are primarily interested in involutive commutative quantales, but note 
every 
commutative quantale can be equipped with the trivial involution. A 
quantale has a least element $\bot$, defined to be the join of the 
empty set, and this is an absorbing element, i.e. for all $x \in Q$ we 
have $x \cdot \bot = \bot$. We assume all quantales are non--trivial, that 
is, $\bot 
\not=\top$, where $\top = \bigvee\limits_{x \in Q} x$.

\begin{remark}\label{rem:QuantalIsSemiring}
An involutive quantale $Q$ is a $*$--semiring with addition given by the join 
and multiplication 
given by the monoid operation. The bottom element 
$\bot$ is the zero element of the semiring and will hence be denoted $0$. We 
say a quantale is 
\emph{zero--divisor free} if it is zero--divisor free as a semiring.
\end{remark}

\begin{definition}
For a commutative involutive quantale $Q$, the category of 
\emph{quantale--valued 
relations} $\RelQ$ has sets as objects and morphisms $f:X \to Y$ consist of 
functions $f: X\times Y \to Q$. For $f:X \to Y$ and $g :Y \to Z$ composition 
is defined where $g\circ f: X \times Z \to Q$ by $g\circ f(x,z) = 
\bigvee\limits_{y \in Y} f(x,y)\cdot g(y,z)$.
We say that a morphism $f:X \to Y$ in $\RelQ$ 
\emph{relates} $x\in X$ to $y \in Y$ if $f(x,y) \not=0$. 
\end{definition}

The category $\RelQ$ is a
$\dg$--symmetric monoidal category with $\dg$--biproducts. The monoidal 
product is given by the cartesian product, with unit the 
one element set, the biproduct is given by 
disjoint union, and the dagger is given by reordering and pointwise application 
of the involution $f^\dg(y,x) = f(x,y)^*$.
\begin{example}
Any complete Heyting algebra or Boolean algebra is a quantale. In particular 
the two--element Boolean algebra $\textbf{2}= \{0,1\}$, where the corresponding 
category $\RelB$ is the usual the category of sets and relations $\Rel$. The 
intervals $[0,1]$ and $[0,\infty]$ are quantales when equipped with the usual 
multiplication, and where $\bigvee\limits S= \sup \, S$.
\end{example}

We now turn our attention to the category $\RelQ\Algvn(X)$ for a set $X$. 
Clearly the scalars $\Hom(I,I)\cong Q$, and for each set $X$ the 
$Q$--semialgebra  $\Hom(X,X)$ is a quantale, with the join given pointwise 
and multiplication given by morphism composition.

\begin{lemma}\label{thm:IsSubquantale}
Each $\AlgA$ in $\RelQ\Algvn(X)$ is a subquantale of $\Hom(X,X)$.
\proof{ By definition $\AlgA$ is a subsemiring, we need to show that $\AlgA$ is 
closed under arbitrary joins. Let $B \subset \AlgA$ be any subset, we need to 
show that $\bigvee\limits_{x 
\in B} x \in \AlgA$. Let $g \in \AlgA'$ then for all $x \in B$ we have $g\cdot 
x= x 
\cdot g$. So we have $g \cdot \bigl( \bigvee\limits_{x \in B} x \bigr)= 
\bigvee\limits_{x \in B} (g \cdot x) = \bigvee\limits_{x \in B} (x \cdot g) = 
\bigl( \bigvee\limits_{x \in B} x \bigr) \cdot g$ and hence $\bigvee\limits_{x 
\in B} x \in \AlgA''$, and since $\AlgA$ is von Neumann $\bigvee\limits_{x 
\in B} x \in \AlgA$, as required.
\QED}
\end{lemma}

We now give an important structure theorem for these semialgebras.

\begin{theorem}\label{thm:CompletelyDecomposable}
Let $\Quant$ be a commutative ZDF quantale and let $\AlgA \in 
\RelQ\Algvn(X)$. There are primitive subunital idempotents $\{ e_i \}$ such 
that $\AlgA = \prod\limits_{i} e_i \AlgA$, a direct product of 
$S^*$--semialgebras.

\proof{ Let $f:X \to X$ be a $Q$--relation. Let $\supp(f)\subset X$, the 
\emph{support} of $f$ be the set of elements $x$ such that there exits $y\in X$ 
such that $f$ relates $x$ to $y$. Let $\cosupp(f)\subset X$, the 
\emph{cosupport} of $f$ be the set of elements $x$ such that there exists $y\in 
X$ such that $f$ relates $y$ to $x$. First, we 
claim 
that for $Q$--relations satisfying $f \circ f^\dg = f^\dg \circ f$ we have 
$\supp(f)= \cosupp(f)$. Suppose $x\in \supp(f)$ then if $Q$ is ZDF then $f^\dg 
\circ f$ relates $x$ to itself. However if $x \not\in \cosupp(f)$ then clearly 
$f \circ f^\dg$ cannot relate $x$ to any other element and hence $x\in \supp(f)$ 
iff $x \in \cosupp(f)$. So $X = \supp(f) \sqcup \overline{\supp(f)}$ and $f$ 
has a corresponding matrix representation 
$f = \bigl(\begin{smallmatrix}
        f_1 & 0 \\
        0 & 0 \\
       \end{smallmatrix}\bigr)$.
For each $f\in \AlgA$ let $f_\supp = \bigl(\begin{smallmatrix}
        \id{} & 0 \\
        0 & 0 \\
       \end{smallmatrix}\bigr) $ be the relation which is the identity 
on 
the support of $f$ and zero otherwise. 

Let $g = \bigl(\begin{smallmatrix}
        g_1 & g_2 \\
        g_3 & g_4 \\
       \end{smallmatrix}\bigr)\in \AlgA'$ then in particular $g\circ f = f \circ 
g$ and hence $\bigl(\begin{smallmatrix}
        g_1 f_1 & 0 \\
        g_3 f_1 & 0 \\
       \end{smallmatrix}\bigr) = \bigl(\begin{smallmatrix}
        f_1 g_1 & f_2 g_2 \\
        0 & 0 \\
       \end{smallmatrix}\bigr)$ and so if $Q$ is ZDF then $g_2= 0$ and $g_3 = 
0$, and hence $g = \bigl(\begin{smallmatrix}
        g_1 & 0 \\
        0 & g_4 \\
       \end{smallmatrix}\bigr)$. Then clearly  $\bigl(\begin{smallmatrix}
        g_1 & 0 \\
        0 & g_4 \\
       \end{smallmatrix}\bigr) \bigl(\begin{smallmatrix}
        1 & 0 \\
        0 & 0 \\
       \end{smallmatrix}\bigr) =\bigl(\begin{smallmatrix}
        1 & 0 \\
        0 & 0 \\
       \end{smallmatrix}\bigr) \bigl(\begin{smallmatrix}
        g_1 & 0 \\
        0 & g_4 \\
       \end{smallmatrix}\bigr)$, i.e. $f_\supp \circ g = g \circ f_\supp$, and 
hence we have shown that $f_\supp \in \AlgA''$, and hence by the assumption 
that $\AlgA$ is von Neumann we have $f_\supp \in \AlgA$. By a similar argument 
$f_{\overline{\supp}} = \bigl(\begin{smallmatrix}
        0 & 0 \\
        0 & \id{} \\
       \end{smallmatrix}\bigr)$ also belongs to $\AlgA$.

Consider the collection of elements $f_\supp$ for all $f \in \AlgA$. Each 
$f_\supp$ corresponds with a subset of $X$ and hence this collection forms a 
Boolean subalgebra 
of $P(X)$, the powerset of $X$. By Lemma \ref{thm:IsSubquantale} $\AlgA$ has 
all joins and hence this collection 
of subunital maps forms a complete Boolean subalgebra of $P(X)$ which by
\cite[Chap. 14, Theorem 8]{GivantHalmos2009:BooleanAlgebras} is atomic. The 
atoms $e_i$ of this Boolean algebra are the primitive subunital idempotents of 
$\AlgA$, and $1_\AlgA = 
\bigvee e_i$. For every 
element $f \in \AlgA$ we have $f = \bigvee f\circ e_i$ for pairwise orthogonal 
subunital idempotents, and hence $\AlgA$ is the direct product of the 
subalgebras $e_i \AlgA$.
\QED}
\end{theorem}

For a commutative ZDF quantale $Q$ and for any set $X$ Theorem 
\ref{thm:CompletelyDecomposable} states that all
semialgebras in $\RelQ\Algvn(X)$ are \emph{completely 
decomposable}, that is, a direct sum of their indecomposable 
(non--unital) subalgebras $e_i \AlgA \subset \AlgA$. We 
call each $e_i \AlgA$ for $e_i$ a primitive subunital idempotent an 
\emph{indecomposable component} of $\AlgA$.

We now give a characterisation of the prime spectrum for semialgebras of 
quantale--valued relations.

\begin{lemma}\label{lem:CharacterisingPSpec}
For $Q$ a commutative ZDF quantale and $\AlgA$ an 
object in
$\RelQ\Algvn(X)$, then $J \subset \AlgA$ is a $k^*$--prime ideal iff if it the 
kernel 
of some $S^*$--semialgebra homomorphism $\gamma: \AlgA \to \AlgBool$.
\end{lemma}

\begin{lemma}\label{lem:SpecAndPrimitiveIdempotents}
For each semialgebra homomorphism $\gamma:\AlgA\to \AlgBool$ there is 
exactly one primitive subunital idempotent $e_a$ in $\AlgA$ such that 
$\gamma(e_a) = 1$.

\proof{First we show there is at most one primitive idempotent $e_a$ such that 
$\gamma(e_a)= 
1$. Suppose there is 
another $e_b$ such that $\gamma(e_b)= 1$. Since $e_a$ and $e_b$ are 
orthogonal we have $\gamma(e_a) \gamma(e_b) \not= \gamma(e_a \circ e_b)$, and 
hence there is at most one $e_a$ such that $\gamma(e_a) = 1$. Suppose however 
there are no primitive idempotents which map to 1. We still have 
$\gamma(1_\AlgA) = \gamma(\bigvee\limits{e_i}) = 1$. If there is only a finite 
number 
of primitive idempotents then we have $\gamma(\bigvee\limits{e_i}) = 
\bigvee\gamma(e_i)$, a contradiction, and hence there is exactly one primitive 
idempotent satisfying $\gamma(e_a) = 1$. Suppose then that there are an 
infinite number of primitive idempotents. Partition the primitive idempotents 
into two infinite sets $K$ and $L$. Then $\gamma(1_\AlgA) =\gamma(\bigvee e_i) 
= \gamma(\bigvee\limits_K e_k) + \gamma(\bigvee\limits_L e_l)  = 1$ but clearly 
$\gamma(\bigvee\limits_K e_k) \gamma(\bigvee\limits_L e_l)=0$ and hence either 
$\gamma(\bigvee\limits_K e_k)=0$ or $\gamma(\bigvee\limits_L e_l)=0$. Suppose 
$\gamma(\bigvee\limits_L e_l)=0$ then by Lemma \label{lem:CharacterisingPSpec} 
$\ker(\gamma)$ is a prime ideal, however there are elements $e_{k_1}$ and 
$e_{k_2}$ in $K$ and 
therefore not in $\ker(\gamma)$ and hence we have $e_{k_1} \cdot e_{k_2}= 0$ 
contradicting the primeness of $\ker(\gamma)$, and hence for each semialgebra 
homomorphism $\gamma: \AlgA \to \AlgBool$ there is exactly one primitive 
idempotent such that $\gamma(e_a)=1$.
\QED}
\end{lemma}

\begin{theorem}\label{thm:NicePrimeIdeals}
For $Q$ a ZDF quantale and $\AlgA$ in $\RelQ\Algvn(X)$ with decomposition 
$\AlgA = \prod\limits_{i} e_i \AlgA$. For each indecomposable subalgebra $e_a 
\AlgA$ the \emph{complement} $\overline{e_a \AlgA}= \prod\limits_{i \not=a} e_i 
\AlgA$ of $e_a \AlgA$ is a prime ideal.
\proof{This follows directly form Lemma \ref{lem:SpecAndPrimitiveIdempotents}, 
simply define the map $\gamma_a:\AlgA \to \AlgBool$ with kernel $\overline{e_a 
\AlgA}$, sending all other elements to 1.
\QED}
\end{theorem}

Although we will see in Example \ref{example:1} that the prime spectrum and the 
Gelfand spectrum for 
$\RelQ$ do not coincide in general, the following theorem shows that they are 
related.

\begin{lemma}\label{lem:NaturalTransformationsGP}
For $Q$ a commutative ZDF quantale there are natural transformations 
$\xi:\GSpec \to \PSpec$ and $\tau: \PSpec \to \GSpec$ such that $\xi\circ \tau 
\cong \id{}$.

\proof{ For $Q$ a quantale there is exactly one quantale homomorphism 
$!:\AlgBool \to Q$. For $Q$ a ZDF quantale there is at least one homomorphism 
$w :Q \to \AlgBool$, 
which sends all non--zero elements to 1.
Since $\PSpec$ can be characterised by the collection of homomorphisms 
$\gamma: \AlgA \to \AlgBool$ let $\tau(\gamma) = ! \circ \gamma$. Similarly 
for 
$\rho:\AlgA \to Q$ define $\xi(\rho)= w \circ \rho$. Naturality is easy to 
check and clearly $w \circ ! \circ \gamma = \gamma$, as required. \QED}
\end{lemma}

In Sect. \ref{sec:Topologising} we discuss a topological interpretation of 
this map $\xi_\AlgA: \GSpec(\AlgA) \to \PSpec(\AlgA)$, in particular how to 
relate the prime spectrum to the state space interpretation of the Gelfand 
spectrum of Figure 1.

The following theorem follows directly from Lemma 
\ref{lem:NaturalTransformationsGP}.
\begin{theorem}\label{thm:GSpecAndPSpec}
For a ZDF quantale $Q$, the Gelfand spectrum for $\RelQ\Algvn(X)$ has a global 
section if and only if the prime spectrum has a global section.
\end{theorem}

\begin{example}\label{example:1}
Let $Q$ be the commutative involutive quantale $[0,1]$ with usual 
multiplication, trivial involution, and where $\bigvee\limits S= \sup \, S$. 
Let 
$X$ be a two element set and consider $\AlgA$ the von Neumann $Q$--semialgebra
$
\AlgA \, =\, \big\{ \bigl(\begin{smallmatrix}
            p & 0 \\
            0 & q
           \end{smallmatrix}\bigr) \ | \ p,q \in Q \ \big\}\, \cong \, Q \oplus 
Q$
It is easy to see that there are four elements of $\PSpec(\AlgA)$:
\[
J_1 = \big\{ \bigl(\begin{smallmatrix}
            p & 0 \\
            0 & 0
           \end{smallmatrix}\bigr) \ | \ p \in Q \ \big\} \qquad \qquad J_2 = 
\big\{ 
\bigl(\begin{smallmatrix}
            p & 0 \\
            0 & q
           \end{smallmatrix}\bigr) \ | \ p \in Q, \ q < 1 \ \big\}
\]\[
K_1 = \big\{ \bigl( \begin{smallmatrix}
            0 & 0 \\
            0 & q
           \end{smallmatrix}\bigr) \ | \ q \in Q \ \big\} \qquad \qquad K_2 = 
\big\{ 
\bigl(\begin{smallmatrix}
            p & 0 \\
            0 & q
           \end{smallmatrix}\bigr) \ | \ q \in Q, \ p < 1 \ \big\}
\]

There are three semialgebra homomorphisms from $Q$ to itself: $u:Q \to Q$ 
defined $u(x) = 1$ for all 
$x\not=0$;  $d:Q \to Q$ defined  $d(x)=0$ for all $x<1$; and the 
identity 
$\id{}:Q \to Q$. Hence there are six homomorphisms
\[
 \varphi_1= \langle d ,0\rangle: Q \oplus Q 
\to Q \quad\quad   \varphi_2 = \langle u,0\rangle: Q \oplus Q 
\to Q \quad\quad   \varphi_3= \langle \id{},0\rangle: Q \oplus Q 
\to Q
\]
\[
 \theta_1= \langle 0,d \rangle: Q \oplus Q 
\to Q \quad \quad   \theta_2 = \langle 0, u \rangle: Q \oplus Q 
\to Q \quad \quad   \theta_3= \langle 0,\id{} \rangle: Q \oplus Q 
\to Q
\]
corresponding to the six elements of $\GSpec(\AlgA)$.
\end{example}

\section{A Proof of Non--Contextuality}

We now show that for a ZDF quantale $Q$ every object $X$ in $\RelQ$ is 
Kochen--Specker non--contextual. We do this by showing that picking an 
element from the underlying set $X$ allows one to construct a global section 
of $\PSpec$. By Theorem 
\ref{thm:GSpecAndPSpec} we can then conclude that $\GSpec$ has global 
sections and thus every object 
in $\RelQ$ is Kochen--Specker non--contextual. We then show a partial converse 
to this result, that every global section for $\PSpec$ in turn picks out an 
element from $X$.

\begin{theorem}\label{thm:MainTheorem}
For $Q$ a commutative ZDF quantale, and $X$ a set, each $x \in X$ 
determines a global section of the prime spectrum $\PSpec: \RelQ\Algvn(X)^{\op} 
\to \Set$.
\proof{ We show that each element $x \in X$ determines a global section. By 
Theorem \ref{thm:CompletelyDecomposable} each semialgebra $\AlgA$ in 
$\RelQ\Algvn(X)$ has a decomposition $\prod\limits_i e_i \AlgA$ for subunital 
idempotents $e_i$. Note that $x$ is in the support of exactly one of the 
primitive subunital idempotents, which we will denote $e_x$.
By Theorem \ref{thm:NicePrimeIdeals} $\overline{e_x \AlgA}$ is a prime ideal. 
Let $\tilde{x}_\AlgA: \AlgA \to \AlgBool$ be 
the map corresponding to this prime ideal defined $\tilde{x}_\AlgA(e_x)= 1$. 
The claim is that $\tilde{x}$ determines a natural transformation. We 
need to show that for each $\AlgA \hookrightarrow \AlgB$ that the restriction 
of 
$\tilde{x}_\AlgB$ so $\AlgA$ is equal to $\tilde{x}_\AlgA$. Let $\AlgB = 
\prod\limits_j d_j \AlgB$ with $\tilde{x}_\AlgB (d_x)=1$. Since 
$e_x$ and $d_x$ both relate $x$ to itself we have $e_x \circ d_x \not= 
0$. Clearly then $e_x \circ d_x$ is a non--zero element of the subsemialgebra 
$d_x \AlgB \subset \AlgB$ and hence $\tilde{x}(e_x \circ d_x) = 1$. This 
implies that $\tilde{x}_\AlgB(e_x) = 1$ and therefore $\tilde{x}_\AlgA (e_x)= 
\tilde{x}_\AlgB (e_x)$.
\QED}
\end{theorem}

Central to the proof of Theorem \ref{thm:MainTheorem} is reducing the problem 
to consider the partitions of the underlying set $X$. The proof of the 
Kochen--Specker theorem 
also reduces the problem to a consideration of the ``partitions'' on the 
Hilbert space $H$, that is, the orthonormal bases of $H$. At the heart 
of the difference between the contextuality results for $\Hilb$ and $\RelQ$ is 
that given an element of a set $X$ we can pick a component from every partition 
of $X$ in a canonical way. However, for a Hilbert space if we choose a vector 
$| \psi \rangle \in H$ there is not a canonical way of picking an element from 
each orthonormal basis of $H$.

This non--contextuality result for $\RelQ$ is consistent with a theorem which 
states 
the category of finite sets and relations is Mermin--local 
\cite{GogiosoZeng2015:MerminNonlocality}, lending some credibility to our 
definition of
Kochen--Specker contextuality.
We now show a partial converse of Theorem \ref{thm:MainTheorem}, that 
is, 
every global section of $\PSpec$ isolates some $x \in X$, although we do not 
claim that every global section is of the form $\tilde{x}$ as defined in the 
proof of 
Theorem \ref{thm:MainTheorem}.
\begin{lemma}\label{lem:TrivialIsVonNeumann}
For $X$ a set, the set of relations $\AlgE = \{ \ q \bullet \id{X}:X \to X \ | 
\ q \in Q \ \}$ belongs to $\RelQ\Algvn(X)$.
\end{lemma}

We call $\AlgE$ (as defined in Lemma \ref{lem:TrivialIsVonNeumann}) the 
\emph{trivial semialgebra on} $X$. Clearly there is an inclusion $\AlgE 
\hookrightarrow \AlgA$ for every $\AlgA$ in $\RelQ\Algvn(X)$. 

\begin{lemma}\label{lem:ClosedUnderJoins}
Suppose $\AlgA \subset \Hom(A,A)$ belongs to $\RelQ\Algvn(A)$ and suppose 
$\AlgB \subset \Hom(B,B)$ belongs to $\RelQ\Algvn(B)$ then $\AlgA \oplus \AlgB 
\subset \Hom(A\sqcup B, A \sqcup B)$  belongs to $\RelQ\Algvn(A\sqcup B)$.
\end{lemma}

\begin{lemma}\label{lem:ClosedUnderRestriction}
If $\AlgA = e_1\AlgA \oplus e_2\AlgA$ belongs to $\RelQ\Algvn(A)$ where $e_1$ 
is the identity morphism 
on some subset $E \subset A$ then $e_1\AlgA$ viewed as a subsemialgebra 
$e_1\AlgA \subset \Hom(E,E)$ belongs to $\RelQ\Algvn(E)$.
\end{lemma}

\begin{theorem} 
Let $Q$ be a ZDF quantale and $X$ an object in $\RelQ$. Every global section 
$\chi:T \to \PSpec(-)$ uniquely determines some $x \in X$.

\proof{ By Lemma \ref{lem:SpecAndPrimitiveIdempotents} for $\AlgA = 
\prod\limits_i e_i \AlgA$ there is one primitive idempotent element $e_a$ such 
that $\chi(e_a) = 1$. For $\AlgB = \prod\limits_j d_j \AlgB$ there is one $d_b$ 
such that $\chi(d_b)=1$. We claim that for $e_a$ and $d_b$ we have $e_a \circ 
d_b\not= 0$.

Let $E_a = \supp(e_a)$ and $E_b = \supp(e_b)$ and let
$\AlgE_1$ be the 
trivial semialgebra defined on the set $X \backslash (E_a \sqcup E_b)$. Let 
$\AlgE_2$ be the trivial semialgebra on $X\backslash E_a$ and $\AlgE_3$ be the 
trivial semialgebra on $X\backslash E_b$. Hence we have unital subsemialgebra 
inclusions 
\[
\begin{tikzpicture}
\node(A) at (0,0) {$e_a \AlgA \oplus \AlgE_2$};
\node(B) at (-1,1) {$\AlgA$};
\node(C) at (1,1) {$e_a \AlgA \oplus d_b \AlgB \oplus \AlgE_1$};
\node(E) at (2,0) {$d_b \AlgB \oplus \AlgE_3$};
\node(F) at (3,1) {$\AlgB$};
\draw[left hook->] (A) to node {}(B);
\draw[right hook->] (A) to node {}(C);
\draw[left hook->] (E) to node {}(C);
\draw[right hook->] (E) to node {}(F);

\end{tikzpicture}
\]
By naturality, if $\chi_\AlgA(e_a)=1$ then $\chi_{e_a \AlgA \oplus e_b 
\AlgB \oplus \AlgE_1}(e_a)=1$ which implies that $\chi_{e_a \AlgA 
\oplus e_b 
\AlgB \oplus \AlgE_1}(e_b)=0$, which in turn implies that $\chi_\AlgB(e_b)=0$, 
which 
is a contradiction.
Since there is an algebra $\AlgA = \prod\limits_{x \in X} Q$, picking a 
global section for this algebra amounts to picking a singleton from $X$.
\QED}
\end{theorem}

\begin{remark}
Spekkens toy theory \cite{Spekkens2007:Epistemic} can be modelled in $\Rel$ 
using Frobenius algebras as a notion of observable 
\cite{CoeckeEdwards2008:ToyQuantum}, and hence by Remark 
\ref{rem:FrobeniusLifts} can be modelled by commutative von Neumann 
semialgebras. In 
Spekkens Toy theory the \emph{ontic states} of 
the physical system, which represent local hidden variables, are 
represented by the singleton elements of the underlying set, and hence we see a 
correspondence between the ontic states of the theory and the global sections 
in the model.
\end{remark}

\section{Topologising the State Space}
\label{sec:Topologising}

The concept of the ``spectrum'' of an algebraic object is a broad one, 
appearing across many fields of 
mathematics: it lies at the heart of a family of deep 
results connecting algebra and topology \cite{Johnstone1982:StoneSpaces}; it 
is a fundamental concept in algebraic geometry 
\cite{Smith2014:AlgebraicGeometry}; and it is central to the algebraic 
approach to classical physics described in Figure 1. 
\cite{Nestruev2003:SmoothManifoldsAndObservables}. In each case one endows the 
spectrum of an algebraic object with a topology called the \emph{Zariski 
topology}. Here we
extend the definition of Zariski topology to the $k^*$--ideals of a semialgebra 
and to the characters 
on a semialgebra and hence to the prime spectrum and Gelfand spectrum of 
$S^*$--semialgebras.

\begin{definition}\label{def:ZariskiTopology} 
Let $\catA$ be a locally small $\dg$--symmetric monoidal category with finite 
$\dg$--biproducts.  Let $X$ be some object, and let $\AlgA$ be an object in 
$\catA\Algvn(X)$.  For each ideal $J\subset \AlgA$ define the sets 
$\mathbb{V}_P(J) = \{ K \in \PSpec(\AlgA) \ | \ 
J \subset K \ 
\}$. We take the collection of $\mathbb{V}_P(J)$ to be a basis of closed sets 
for 
the \emph{Zariski topology} on $\PSpec(\AlgA)$.
Consider the set $\GSpec(\AlgA)$. For each ideal 
$J\subset \AlgA$ define the set $\mathbb{V}_G(J) = \{ \rho \in \Spec(\AlgA) \ | 
\ 
J 
\subset \ker(\rho) \ \}$. We take the collection of $\mathbb{V}_G(J)$ to be a 
basis of 
closed sets for the \emph{Zariski topology} on $\GSpec(\AlgA)$.
\end{definition}

Hence, under the interpretation of Figure 
1. we see that our sets of states are in fact topological spaces. Recall, a 
space is $T_0$ if all points are \emph{topologically 
distinguishable}, 
that is, for every pair of points $x$ and $y$ there is at least one open set 
containing one but not both of these points.

\begin{theorem}\label{thm:FunctorToTop}
For an $S^*$--semialgebra $\AlgA$ the Zariski topology on $\PSpec(\AlgA)$ is  
compact and $T_0$, and for $i:\AlgA \hookrightarrow \AlgB$ the 
function 
$\tilde{i}: \PSpec(\AlgB) \to \PSpec(\AlgA)$ in continuous with respect to this 
topology.
For an $S^*$--semialgebra $\AlgA$ the Zariski topology on $\GSpec(\AlgA)$ is  
compact, and for $i:\AlgA \hookrightarrow \AlgB$ the function 
$\tilde{i}: \GSpec(\AlgB) \to \GSpec(\AlgA)$ in continuous with respect to this 
topology.
\end{theorem}
Theorem \ref{thm:FunctorToTop} states the the prime spectrum and Gelfand 
spectrum give us functors of the form
\[
\begin{tikzpicture}
\node(A) at (0,0) {$\catA\Alg(A)^{\op}$};
\node(B) at (3,0) {$\CTop$};
\draw[->](A) to node [above]{$\PSpec$}(B);
\end{tikzpicture}\qquad\qquad
\begin{tikzpicture}
\node(A) at (0,0) {$\catA\Alg(A)^{\op}$};
\node(B) at (3,0) {$\CTop$};
\draw[->](A) to node [above]{$\GSpec$}(B);
\end{tikzpicture}
\]

Note the Gelfand 
spectrum need not even be $T_0$ in general, as we will see in 
Example \ref{example:2}.

Theorem \ref{thm:GSpecAndPSpec} relates the prime spectrum and the Gelfand 
spectrum for the case when $\catA$ is the category $\RelQ$ for a $Q$ a ZDF 
quantale. The following theorem gives us an insight into the 
nature of this relationship in terms of the topological structure on these 
spectra.

\begin{theorem}\label{thm:GSpecAndPSpecAsTopologcalSpaces}
For $Q$ a ZDF quantale and $\AlgA$ in $\RelQ\Algvn(X)$, each $\xi_\AlgA: 
\GSpec(\AlgA) \to \PSpec(\AlgA)$, as defined in Theorem 
\ref{thm:GSpecAndPSpec},
is a quotient 
of topological spaces
where $\rho_1 \sim \rho_2$ iff $\rho_1$ and $\rho_2$ are not distinguishable by 
the Zariski topology.
\end{theorem}

The map $\xi_\AlgA$ identifies those characters which have the same kernel, 
which are precisely those characters which the Zariski topology on 
$\GSpec(\AlgA)$ cannot distinguish.

Theorem \ref{thm:GSpecAndPSpecAsTopologcalSpaces} allows us to think of 
$\PSpec(\AlgA)$ as a coarse--graining of the state space $\GSpec(\AlgA)$ of our 
physical system. To illustrate this we revisit Example \ref{example:1}.

\begin{example}\label{example:2}
Let $\AlgA$ be as in Example \ref{example:1}. The Zariski 
topology on $\PSpec(\AlgA)$ has a basis consisting of the sets $\{ J_1, J_2 
\}$, $\{K_1, K_2\}$, 
$\{J_1\}$, and $\{K_1\}$. It is easy to check that this topology is $T_0$ but 
that it is not $T_1$ 
and therefore not
Hausdorff. For $\GSpec(\AlgA)$ the Zariski topology has a basis of 
closed 
sets $\{ \varphi_1, 
\varphi_2, \varphi_3 \}$, $\{ \varphi_1 \}$, $\{ \theta_1, 
\theta_2, \theta_3 \}$, $\{ \theta_1 \}$. It is easy to check that there is no 
open set distinguishing $\varphi_2$ and $\varphi_3$ from one 
another, nor 
$\theta_2$ from $\theta_3$ as these respective pairs of characters have the 
same kernels and 
hence $\GSpec(\AlgA)$ fails even to be $T_0$. Note that 
$\xi_\AlgA(\varphi_2) =  \xi_\AlgA(\varphi_3) = K_1$ and $\xi_\AlgA(\theta_2) = 
 \xi_\AlgA(\theta_3) = J_1$, and hence the topologically indistinguishable 
points are identified by $\xi_\AlgA$.
\end{example}

\bibliography{all}

\end{document}